\documentclass[a4paper,11pt]{article}
\usepackage{jheppub} 

\usepackage{amsmath,amssymb,bm}
\usepackage{mathtools}
\usepackage{color}
\usepackage{physics}
\usepackage{comment}

\usepackage{colortbl}

\newcommand{\vp}{\varphi}

\newcommand{\p}{\partial}

\newcommand{\cl}[1]{\mathcal{ #1}}
\newcommand{\tx}[1]{\mathrm{ #1}}

\newcommand{\eqn}[1]{\begin{equation}\begin{split}#1\end{split}\end{equation}}

\arxivnumber{} 
\preprint{YITP-24-71}
\title{\boldmath  AdS/CFT correspondence for the $O(N)$ invariant critical $\vp^4$ model in 3-dimensions by the conformal smearing  }

\author[a]{Sinya Aoki}
\author[b]{Kiyoharu Kawana}
\author[a]{Kengo Shimada}
\affiliation[a]{Center for Gravitational Physics and Quantum Information, Yukawa Institute for Theoretical Physics, Kyoto University,\\
Kitashirakawa Oiwakecho, Sakyo-Ku, Kyoto 606-8502, Japan}
\affiliation[b]{School of Physics, Korea Institute for Advanced Study,\\
Seoul 02455, Korea}
\emailAdd{saoki@yukawa.kyoto-u.ac.jp}
\emailAdd{kkiyoharu@kias.re.kr}
\emailAdd{kengo.shimada@yukawa.kyoto-u.ac.jp}

\abstract{
We investigate a structure of a 4-dimensional bulk space constructed from the $O(N)$ invariant critical $\vp^4$ model in 3-dimension using the conformal smearing. We calculate a bulk metric corresponding to the information metric and the bulk-to-boundary propagator for a composite scalar field $\vp^2$
in the large $N$ expansion.
We show that the bulk metric describes an asymptotic AdS space at both UV (near boundary) and IR (deep in the bulk) limits, which correspond to the asymptotic free UV fixed point and the Wilson-Fisher IR fixed point of the 3-dimensional $\vp^4$  model, respectively. 
The bulk-to-boundary scalar propagator, on the other hand, encodes $\Delta_{\vp^2}$ (the conformal dimension of $\vp^2$) into its $z$ (a coordinate in the extra direction of the AdS space) dependence. 
Namely it correctly reproduces not only $\Delta_{\vp^2}=1$ at UV fixed point but also $\Delta_{\vp^2}=2$
at the IR fixed point for the boundary theory. Moreover, we confirm consistency with the GKP-Witten relation in the interacting theory that the coefficient of the $z^{\Delta_{\vp^2}}$ term in $z\to 0$ limit agrees exactly with the two-point function of $\vp^2$ including an effect of the $\vp^4$ interaction.   }

\begin{document}
\maketitle
\flushbottom


\section{Introduction}
The AdS/CFT correspondence~\cite{Maldacena:1997re,Gubser:1998bc,Witten:1998qj} plays a crucial role in understanding the holographic nature of gravity and may give a hint on its quantization, but it is still mysterious even though plenty of evidences and applications exist after the first proposal.
While a large part of the AdS/CFT correspondence can be understood 
in the context of the closed string/open string duality with $D$ branes, a complete understanding of this duality has not been attained yet.
It is widely accepted, however, that a particular type of CFT can be holographic dual to bulk gravity theories, and such theories are called holographic CFT.  

On the other hand, it is generally believed that the AdS radial direction emerges as the energy scale of 
a renormalization group transformation applied to the dual CFT at the boundary. (For example, see Ref.~\cite{Heemskerk:2010hk}).
Among studies in such a direction, 
the continuum version of multi-scale entanglement renormalization ansatz (cMERA)
has been employed as a real space quantum renormalization group, in order to generate the extra dimension
as a level of the coarse-graining~\cite{Nozaki:2012zj,Miyaji:2015yva,Miyaji:2015fia}.
In their approach, the geometric structure of the bulk spacetime at a given time slice is 
determined from a quantum information metric for a CFT state on the equal-time (boundary) surface. 
Indeed the bulk space-time becomes AdS for the vacuum state, and thus the AdS spacetime naturally emerges from the boundary CFT.
More interestingly, their bulk construction can be applied 
not only to holographic CFT but also to generic CFT or even to non-conformal field theories. 

Following the philosophy of Refs.~\cite{Nozaki:2012zj,Miyaji:2015yva,Miyaji:2015fia},
one of the present authors has proposed a similar but different method to construct the Euclidean AdS space from Euclidean CFT by employing a different coarse-graining technique called flow equations and applied it to various cases~\cite{Aoki:2015dla,Aoki:2016env,Aoki:2016ohw,Aoki:2017bru,Aoki:2017uce,Aoki:2018dmc,Aoki:2019bfb,Aoki:2019dim,Aoki:2020ztd}.
Recently, an improved version of the flow equations has been found~\cite{Aoki:2022lye}, and we call the corresponding method a conformal smearing 
since the conformal transformations applied to CFT fields on the boundary are literally mapped to a part of the general coordinate transformations of the smeared fields in the bulk, which is the full isometry of the AdS space. 

In this paper, in order to obtain deeper understanding of a mechanism for an emergent extra dimension in AdS/CFT correspondences, we apply the conformal smearing to the $O(N)$ invariant critical $\lambda\varphi^4$ model in 3-dimensions, which has the asymptotic free UV fixed point and the Wilson-Fisher IR fixed point, where $\lambda$ has mass dimension one and breaks conformal symmetry.
This model is thought to be dual to higher spin theories~\cite{Klebanov:2002ja}, and 
this duality has been investigated in terms of a bi-local field, where a magnitude of its relative coordinate is interpreted as an extra bulk dimension~\cite{Das:2003vw,deMelloKoch:2010wdf,deMelloKoch:2018ivk, Aharony:2020omh}.
The conformal smearing approach is different from theirs. In particular, while 
the bulk geometry is assumed to be AdS in their approach, it is determined by the information metric in our approach.    
Indeed, the previous study employing the Gaussian smearing~\cite{Aoki:2016env}  
has shown that the bulk space becomes the asymptotic AdS space at both UV limit(near boundary) and IR limit (deep in the bulk), whose AdS radii are different in two limits.
Since the Gaussian smearing keeps only a part of the relation between the conformal transformation and the AdS isometry~\cite{Aoki:2017bru}, however, 
results in Ref.~\cite{Aoki:2016env} are insufficient to understand the duality between the bulk theory and the boundary CFT. For example, inequality for the AdS radii between two limits can not be determined, and a change of the conformal dimension of the composite scalar operator between two limits has not been investigated in terms of  the bulk language. 
We, therefore, investigate this duality using the conformal smearing in this paper.

We here summarize the main results of our study.
\begin{itemize}
\item[(1)] The bulk space constructed from the interacting $O(N)$ invariant critical $\varphi^4$ model at $d=3$ by the conformal smearing is the 4-dimensional AdS space at the leading order (LO) in the large $N$ expansion. While, at the next-to-leading order (NLO), 
it becomes the asymptotic AdS space both in the UV and the IR limit, which correspond to the asymptotic free UV fixed point and the Wilson-Fisher IR fixed point, respectively. 
We also observe that the AdS radius increases from UV to IR as $R_{\rm AdS}^{\rm UV} < R_{\rm AdS}^{\rm IR}$ at the NLO.
Our naive interpretation is as follows. Since the number of dynamical degrees of freedom decreases from UV to IR by the renormalization group,  
the corresponding bulk space becomes less "AdS", meaning that the negative cosmological constant reduces in magnitude and gets closer to the flat space. 
Therefore the AdS radius increases. This result, however, is opposite to the prediction of the F-theorem that $R_{\rm AdS}^{\rm UV} > R_{\rm AdS}^{\rm IR}$ \cite{Freedman:1999gp, Myers:2010xs, Myers:2010tj, Jafferis:2011zi, Klebanov:2011gs,Pufu:2016zxm}. 
\item[(2)] The bulk-to-boundary propagator of the $O(N)$ invariant scalar field has been calculated. In the $z\to 0$ limit where $z$ is the radial coordinate of the bulk space, this propagator behaves as $z^1$, showing that the conformal dimension of the corresponding $O(N)$ invariant scalar operator at the boundary is one~\cite{Banks:1998dd}. This values agrees correctly with $\Delta_{\varphi^2}=1$, 
the conformal dimension of the composite scalar operator $\varphi^2$ (the spin zero "current" $J$ in Ref.~\cite{Klebanov:2002ja}) at the UV fixed point.
On the other hand, in the IR limit ($z\to \infty$), the bulk-to-boundary propagator behaves as
\begin{equation}
z^{-2} \simeq \left(\frac{z}{z^2+x^2}\right)^2,
\end{equation}
where $x$ is the boundary coordinates. This behavior corresponds to $\Delta_{\varphi^2}=2$,
which is the conformal dimension of $\varphi^2$ at the IR fixed point.
\end{itemize}
These two findings show that the non-trivial dynamics of the boundary theory generated by the non-conformal interaction term is correctly encoded in the bulk geometry and dynamics.

\section{Model and Conformal smearing}
Throughout the paper, we are working on $d=3$, where $d$ is the dimension of the boundary.  

\subsection{$O(N)$ model in 3 dimensions}
We consider an $O(N)$ invariant model for scalar fields, whose action is given by
\begin{equation}
S(\varphi) = N \int d^3x\left[{Z_\varphi\over 2} \partial^\mu\varphi\cdot \partial_\mu\varphi
+{m^2 Z_m \over 2 }Z_\varphi \varphi\cdot\varphi +{\lambda Z_\lambda\over 4} Z_\varphi^2 ( \varphi\cdot\varphi)^2
\right],  
\label{eq:action}
\end{equation}
where $\varphi\cdot\varphi :=\sum_{a=1}^N \varphi^a\varphi^a$ with  $a$ being an $O(N)$ index, and
$Z_\varphi$, $Z_m$ and $Z_\lambda$  are renormalization constants which relate bare to renormalized quantities as $\varphi_0^a =\sqrt{Z_\varphi}\varphi^a$, $m_0^2 = m^2 Z_m$, and $\lambda_0 = \lambda Z_\lambda$, respectively. 
Note that we have extracted a factor $N$ to consider the large $N$ expansion, and $m^2 Z_m$ in our notation includes the additive mass counter terms.

We calculate correlation functions necessary in this paper in the large $N$ expansion, by employing the Schwinger-Dyson equation in Appendix~\ref{sec:largeN}, and results are summarized below.

In this paper, we consider the critical case,
where the renormalized mass is tuned to be zero.
The 2-pt function $N\langle \varphi^a(x)\varphi^b(y)\rangle :=  \delta^{ab} \Gamma(x-y)$ in this case ($m^2=0$) is given at the NLO in the large $N$ expansion  as
\begin{eqnarray}
\Gamma(x) &=& \int_p  e^{i p x}  \left[ \tilde\Gamma_0(p) +{1\over N} \tilde\Gamma_1(p)\right],\quad \int_p :=  \int {d^3p\over (2\pi)^3},
\end{eqnarray}
where
\begin{eqnarray}
\tilde\Gamma_0(p) &=&{1\over p^2}, \quad 
\tilde\Gamma_1(p) = {X(p^2)\over (p^2)^2}, 
\quad X(p^2) :=\int_Q \left[{1\over (Q-p)^2} -{1\over Q^2}\right] {-2\lambda_0\over 1+\lambda_0 B(Q^2)}
\end{eqnarray}
with $B(p^2)=1/( 8 |p|)$.

The 4-pt function is decomposed as
\begin{eqnarray}
K^{a_1a_2a_3a_4}(x_1,x_2,x_3,x_4) &:=& N^3 \langle \varphi^{a_1}(x_1)\varphi^{a_2}(x_2)\varphi^{a_3}(x_3)\varphi^{a_4}(x_4) \rangle \nonumber \\
&=& \delta^{a_1a_2}\delta^{a_3a_4} K(x_1,x_2;x_3,x_4) + (2\leftrightarrow 3) + (2\leftrightarrow 4), 
\end{eqnarray}
where $K$ at the LO is given by
\begin{eqnarray}
K_0(x_1,x_2;x_3,x_4) &=& \prod_{i=1}^4 \int_{p_i} {e^{i p_i x_i}\over p_i^2} \times (2\pi)^3\delta^{(3)}\left( \sum_{i=1}^4 p_i \right)  { -2\lambda_0 \over 1+\lambda_0 B((p_1 + p_2)^2)}.
\end{eqnarray}

\subsection{Conformal smearing}
In Ref.~\cite{Aoki:2022lye}, the conformal smearing has been introduced to construct bulk field $\phi^a$
from $\varphi^a$ as
\begin{eqnarray}
\phi^a(X):= \int d^3 y\, S(x-y,z) \varphi^a(y)
= \int_p S(p,z) \tilde\varphi^a(p) e^{i p x},
\end{eqnarray}
where $X:=(z,x)$, $\tilde\varphi^a(p)$ and $S(p,z)$ are the Fourier transforms of $\varphi^a(x)$ and $S(x,z)$, respectively, and the smearing kernel in the momentum space is given with the modified Bessel function $K_1$ as
\begin{equation}
    S(p,z) := p z K_1( p z), \quad p:=\vert \vec p\vert~. 
\end{equation}
It is easy to see that $\phi^a(X)$ is the solution to the (conformal) flow equation~\cite{Aoki:2022lye} as
\begin{equation}
    -\eta\partial_\eta^2 \phi^a(X) =\Box_x \phi^a(X), \quad \phi^a(0,x)=\varphi^a(x),\quad \eta:= {z^2\over 4}.
\end{equation}
Furthermore, we define the normalized smeared field as 
\begin{equation}
  \sigma^a(X):= {\phi^a(X) \over \sqrt{\gamma(z)}}, \quad\gamma(z) := \sum_a \langle \phi^a(X)\phi^a(X)\rangle,  
\end{equation}
where $\langle \cdots \rangle$ is a vacuum expectation value in the $O(N)$ model. 
It was shown that the conformal transformations to $\varphi^a(x)$ generate a part of general coordinate transformations applied to the scalar $\sigma^a(X)$~\cite{Aoki:2022lye}.
The translational invariance tells us that $\gamma$ only depends on $z$. 

At the NLO in the large $N$ expansion, we explicitly obtain
\begin{equation}
    \gamma(z) =\gamma_0(z)  +{1\over N}\gamma_1(z),
\end{equation}
where
\begin{equation}
    \gamma_0(z) =\int_p {S^2(p.z)\over p^2}, \quad\gamma_1(z) =\int_p {S^2(p,z) (p^2)^2} X(p^2).
\end{equation}

\section{Bulk metric via the conformal smearing}
In the smearing approach, the bulk metric corresponding to the vacuum state of the boundary theory\footnote{The bulk metric depends on the boundary state. See \cite{Aoki:2023cil} in the case of the metric for the thermal state. } 
can be defined in terms of the normalized smeared field~\cite{Aoki:2015dla} as
\begin{equation}
    g_{AB}(X) :=\ell^2 \langle \partial_A \sigma^a(X) \partial_B \sigma^a(X)  \rangle, 
\end{equation}
which can be interpreted as the Bures (quantum) information metric~\cite{Aoki:2017bru,Aoki:2022lye},
where $\ell$ is some constant of the length scale.  
Note that a similar definition using Bures metric was employed in Ref.~\cite{Nozaki:2012zj}.

Non-zero components of the metric are given at the NLO as
\begin{eqnarray}
    g_{\mu\nu}(z) &=& {\delta_{\mu\nu}\over 3\gamma(z)}\left[F_0(z) +{1\over N} F_1(z)\right], \nonumber \\
    g_{zz}(z) &=& \left({\gamma_z(z)\over 2\gamma(z)}\right)^2 - {\gamma_z(z)\over \gamma^2(z)} \left[G_0(z) +{1\over N} G_1(z)\right]+{1\over \gamma(z)}\left[H_0(z) +{1\over N} H_1(z)\right], 
\end{eqnarray}
where $\gamma_z^{}(z):=\displaystyle {d\gamma(z)\over d z}$ and the LO contributions are expressed as
\begin{eqnarray}
    F_0(z) &=& \int_p S^2(p,z), \quad
    G_0(z) =\int_p {S_z(p,z) S(p.z)\over p^2}, \quad
    H_0(z) =\int_p {S_z(p,z) S_z(p.z)\over p^2}, 
\end{eqnarray}
with $S_z^{}(p,z):=\partial_z^{}S(p,z)$.
On the other hand, NLO corrections are 
\begin{eqnarray}
    F_1(z) &=& \int_p {S^2(p,z)\over p^2} X(p^2), \quad \nonumber \\
    G_1(z) &=& \int_p {S_z(p,z) S(p,z)\over (p^2)^2} X(p^2), \quad 
    H_1(z) = \int_p {S_z(p,z) S_z(p,z)\over (p^2)^2} X(p^2).
\end{eqnarray}

\subsection{Results at the LO}
Using the integration formula~(\ref{eq:Bessel1}),
we obtain 
\begin{eqnarray}
    \gamma_0(z) ={3\over 64 z}, \quad F_0(z) ={45\over 1024 z^3}, \quad G_0(z)=-{3\over 128 z^2}, \quad H_0(z) ={27\over 1024 z^3} .
\end{eqnarray}
Therefore, the metric at the LO is given by
\begin{eqnarray}
    g_{\mu\nu}^{\rm LO}(z) &=& {5 \ell^2\over 16}{\delta_{\mu\nu} \over z^2}, \qquad
    g_{zz}^{\rm LO} = {5 \ell^2\over 16}{1 \over z^2},
\end{eqnarray}
which is the Euclidean AdS metric with the AdS radius~\cite{Aoki:2023cil}
\begin{equation}
    R^{\rm LO}_{\rm AdS}= \ell{\sqrt{5}\over 4} = \ell \sqrt{{\Delta_\varphi (d-\Delta_\varphi)\over d+1}}
\end{equation}
at $d=3$, where $\Delta_\varphi=(d-2)/2$ is the conformal dimension of a free massless scalar.   

\subsection{Results at the NLO}
\begin{figure}[t]
\begin{center}
\includegraphics[width=13cm]{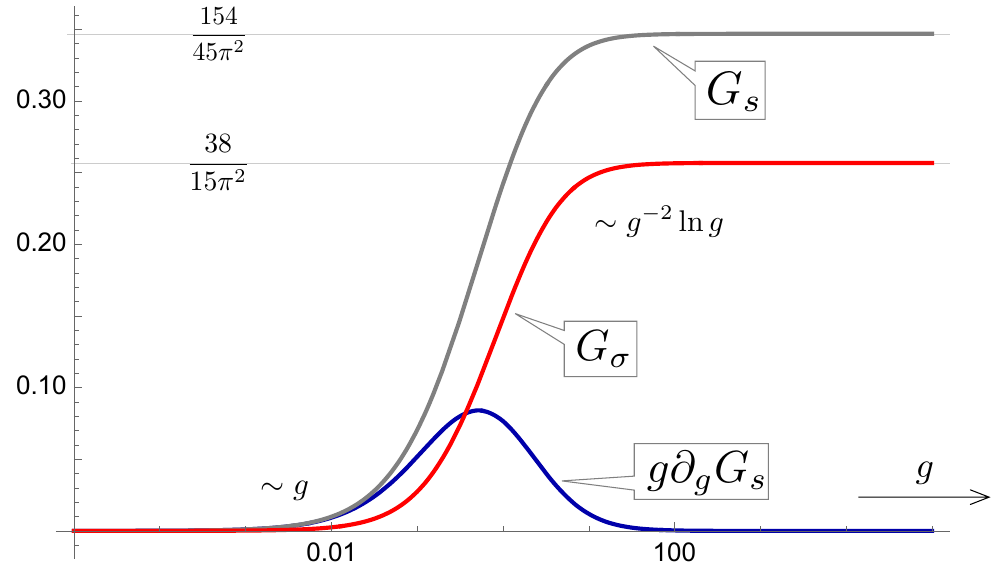}
\end{center}
\caption{The $1/N$ corrections to the metric, $G_s(z)$ (gray) and $G_\sigma(z)$ (red), as a function of $g = \lambda_0 z/8$.
The horizontal axis is on a logarithmic scale for $g$.
Also, $g \p_g G_s = z \p_z G_s$ is depicted for later convenience.}
\label{Fig.Correction_to_AdS_radius}
\end{figure}

The metric at the NLO is given by
\begin{eqnarray}
    g_{\mu\nu}^{\rm NLO}(z) &=& g_{\mu\nu}^{\rm LO}(z) \left[1 +{1\over N} G_s(z)\right], \nonumber \\
    g_{zz}^{\rm NLO}(z) &=& g_{zz}^{\rm LO}(z) \left[1 +{1\over  N}G_\sigma(z)\right],
    \label{eq:metric_NLO}
\end{eqnarray}
where $G_s(z)$ and $G_\sigma(z)$ are defined in Eqs.~(\ref{eq:Gs}) and \eqref{eq:Gr}.
Their behaviors are shown in Fig.\ref{Fig.Correction_to_AdS_radius} as functions of $g=\lambda_0 z /8$.

In the UV limit that $z\rightarrow 0$, from explicit expression of $F_{ij}^n(g)$ in Eq.~(\ref{eq:Fijn_UV}),
we see that the $D^1_{11}$ term in $F_{11}^1(g)$ dominates in the limit. 
Therefore the metric becomes
\begin{eqnarray}
    g_{\mu\nu}^{\rm NLO}(z) &\simeq& g_{\mu\nu}^{\rm LO}(z) \left[ 1 -{g\over N} {128 D^1_{11}\over 3\pi^4}   \right], \nonumber\\
    g_{zz}^{\rm NLO}(z) &\simeq& g_{zz}^{\rm LO}(z) \left[ 1 -{g\over N} {128 D^1_{11}\over 15\pi^4} \right],
\label{eq:metric_UV}
\end{eqnarray}
as $g:=\lambda_0 z/8 \to 0$, where $D^1_{11} = - \pi^2 /4$ is given in Eq.~(\ref{eq:D110}) with $c_1=1$. 
This means that the NLO correction is sub-leading of the order $z$ in the UV limit so that the AdS radius is unchanged: 
$R_{\rm AdS}^{UV}=R_{\rm AdS}^{\rm LO}$ and the metric describes the asymptotic AdS space.

\begin{figure}[t]
\begin{center}
\includegraphics[width=13cm]{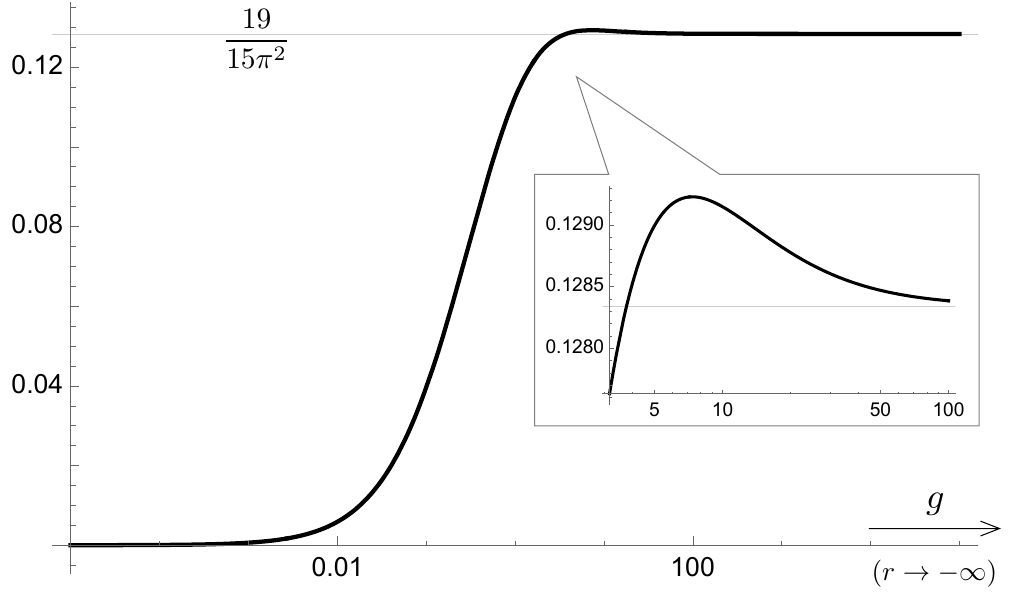}
\end{center}
\caption{The difference between the effective AdS radius and the one in the UV limit (\ref{effective_radius}) is shown as a function of $g=\lambda_0 z/8$ as in Fig.\ref{Fig.Correction_to_AdS_radius}. Here, we have taken the large $N$ limit after multiplying it by $N$.
Its non-monotonicity comes from $z \p_z G_s$, which has the bump as seen in Fig.\ref{Fig.Correction_to_AdS_radius}.
}
\label{Fig.EffectiveRadius}
\end{figure}

In the IR limit that $z\rightarrow \infty$, on the other hand, Eq.~(\ref{eq:GssigIR}) leads to 
\begin{eqnarray}
    g_{\mu\nu}^{\rm NLO} (z) &\simeq& g_{\mu\nu}^{\rm LO} (z)\left[1-{4\over 3\pi^2 N} C_{\rm IR}\right], \nonumber \\
    g_{zz}^{\rm NLO} (z) &\simeq& g_{zz}^{\rm LO} (z)\left[1+{4\over 3\pi^2 N}\left({48\over 5} + 3 C_{\rm IR}\right)\right]
    \label{eq:metric_IR}
\end{eqnarray}
as $g\to\infty$, where
\begin{equation}
    C_{\rm IR} = {512\over 45\pi^2} \int_0^\infty  dp\, p^2\left({15\over 16} -p^2\right)  K_1^2(p) \ln (p^2) = - \frac{77}{30}. 
\end{equation}
By the change of the $z$ coordinate as
\begin{equation}
    \tilde z = z\left[1+{4\over 3\pi^2 N}\left({24\over 5} + 2 C_{\rm IR}\right)\right] ,
\end{equation}
we finally obtain
\begin{eqnarray}
    \tilde{g}_{AB}^{\rm NLO}(\tilde{z})
    &\simeq& (R_{\rm AdS}^{\rm LO})^2 {\delta_{AB}\over \tilde z^2}\left[1+{4\over 3\pi^2 N}\left({48\over 5} + 3 C_{\rm IR}\right)\right],
\end{eqnarray}
which describes the AdS space with the radius given by
\begin{equation}
    R_{\rm AdS}^{\rm IR} := R_{\rm AdS}^{\rm LO} \left[1+{2\over 3\pi^2 N}\left({48\over 5} + 3 C_{\rm IR}\right)\right].
\end{equation}
Since 
\begin{equation}
{48\over 5} + 3 C_{\rm IR} = \frac{19}{10} ,    
\end{equation}
the radius in the IR limit is larger than the one in the UV limit:
$R_{\rm AdS}^{\rm IR} -  R_{\rm AdS}^{\rm UV} = \cl{O}(1/N) >0$.
As an interpolation between UV and IR, we may define the ``effective AdS radius'' by $R^\tx{eff}_\tx{AdS} = 1/A'$ where $A$ is the exponent of the warp factor in the line element written as $ds^2 = e^{2 A}   d x^2  + dr^2$ and the prime is the derivative with respect to $r:= -\int dz \sqrt{g_{zz}}$, whose difference from the AdS radius in the UV limit $R^\tx{UV}_\tx{AdS}$, normalized by $R^\tx{UV}_\tx{AdS}$, is given by
\eqn{
\frac{R^\tx{eff}_\tx{AdS} - R^\tx{UV}_\tx{AdS}}{R^\tx{UV}_\tx{AdS}} = \frac{   G_\sigma + z \partial_z G_s }{2 N} + \cl{O}\qty(N^{-2}) ~. \label{effective_radius}
}
As shown in Fig.\ref{Fig.EffectiveRadius}, this quantity is positive, and increases from $g=0$, the UV limit, to $g\sim 10$. Then, it decreases toward its asymptotic value $19/15 \pi^2$ in the IR limit with $g\to \infty$ corresponding to $r\to - \infty$.
This can be understood as follows. Since the number of effective degrees of freedom decreases from UV to IR by the renormalization group  (smearing in our case), corresponding contributions to the negative cosmological constant in the bulk also decrease, so that the AdS radius increases.
This result, however, is opposite to the prediction by the F-theorem
in quantum field theories at $d=3$~\cite{Jafferis:2011zi, Klebanov:2011gs,Pufu:2016zxm}.
Since the free energy $F$ on a 3-sphere is expected to decrease from UV to IR and the AdS/CFT correspondence predicts $F\propto R^2_{\rm AdS}$, the AdS radius decreases accordingly.
In the pioneering work on the holographic c-theorem \cite{Freedman:1999gp}, it is discussed that the monotonically decreasing behavior of $R^\tx{eff}_\tx{AdS} = 1/A'$ is equivalent to the null energy condition when the bulk theory is the Einstein gravity plus matter fields.
Applied to our model, it means that our bulk theory (analytically continued to the Lorentzian one) should not be the Einstein gravity with the null energy condition always satisfied.
While we do not understand the reason for this discrepancy precisely at this moment, one of the most plausible explanations is that a decrease or increase of the AdS radius is {\it NOT} universal under the renormalization group transformation.
Indeed, in our previous work for the same model by the different smearing \cite{Aoki:2016env}, we have found at the NLO that
$R_{\rm AdS}^{\rm UV} < R_{\rm AdS}^{\rm IR}$ for the Gaussian smearing while $R_{\rm AdS}^{\rm UV} > R_{\rm AdS}^{\rm IR}$ if we add an interaction term to the Gaussian smearing. Therefore it is reasonable to expect that we may realize $R_{\rm AdS}^{\rm UV} > R_{\rm AdS}^{\rm IR}$ by modifying the conformal flow.

\section{Bulk-to-boundary scalar propagator}
In this section, we consider the $O(N)$ invariant scalar field
\eqn{
O(y) := ~  : \varphi^b(y) \varphi^b(y): ~,
}
where $:\ :$ denotes the normal ordering. 
The bulk-to-boundary propagator for this operator can be defined as the correlation function with the corresponding bulk (smeared) operator:
\begin{equation}
    \Pi(X,y) := \left\langle \sigma^a(X) \sigma^a(X) O(y) \right\rangle .  \label{eq:Pi}
\end{equation}

At the LO in the large $N$ expansion, we obtain
\begin{eqnarray}
    \Pi(X,y) &=&{1\over N} \left[ 2 \Pi_0(x-y,z) + \Pi_\lambda(x-y,z)\right] + \cl{O}\left({1\over N^2}\right) \label{eq:Bulk-to-boundary}
\end{eqnarray}
where
\begin{eqnarray}
 \Pi_0(x-y,z) &=& {1\over \gamma_0(z)}\left[\int_p e^{i p (x-y)} S(p,z) \Gamma_0(p)\right]^2\label{eq:Pi0} \\
 &=& 
 {1\over \gamma_0(z)}\left[\int d^3y_1\, S(x-y_1,z) \Gamma_0(y_1-y)\right]^2   
 = {4\over 3\pi^2}\left[ {z\over z^2+(x-y)^2}\right],
\end{eqnarray}
which is nothing but the bulk-to-boundary propagator (except a factor two) in the free theory~\cite{Aoki:2022lye}.
This behavior tells us that the conformal dimension of the $O(N)$ invariant composite scalar operator $\varphi^2$ is given by $\Delta_{\varphi^2}=1$ in the free theory.

On the other hand, the contribution coming from the interaction is given by
\begin{eqnarray}
    \Pi_\lambda(x-y,z) &:=& {1\over \gamma_0(z)}\int d^3y_1\, d^3y_2\,  S(x-y_1,z) S(x-y_2,z) K_0(y_1,y_2;y,y)~,
 \end{eqnarray}
 which is evaluated as
 \begin{eqnarray}
\Pi_\lambda(x-y,z) &=& {1\over \gamma_0(z)}\int_{p_1,p_2}e^{i(p_1+p_2)(x-y)} {S(p_1,z) S(p_2,z)\over p_1^2p_2^2}{-2\lambda_0 B(p_{12}^2)\over 1+\lambda_0 B(p_{12}^2)}
\label{eq:G_UV}\\
    &=& {1\over \gamma_0(z) z^2}\int_{p_1,p_2}{ K_1(p_1) K_1(p_2)\over \vert p_1\vert  \vert p_2\vert} e^{i(p_1+p_2)(x-y)/z}
    {-2g\over p_{12} + g},
    \label{eq:G_IR}
\end{eqnarray}
where $p_{12}:=\vert p_1+p_2\vert$.

Combining (\ref{eq:Pi0}) and (\ref{eq:G_UV}), the LO contribution in Eq.~(\ref{eq:Bulk-to-boundary}) turns out to be
 \begin{eqnarray}
\Pi (x-y, z)&=&
{2\over N \gamma_0(z)}\int_{p_1,p_2}e^{i(p_1+p_2)(x-y)} {S(p_1,z) S(p_2,z)\over p_1^2p_2^2}{1 \over 1+\lambda_0 B(p_{12}^2)}  \nonumber \\
&=& {2\over N \gamma_0(z) z^2}\int_{p_1,p_2} \frac{ K_1(p_1) K_1(p_2)}{ \vert p_1\vert  \vert p_2\vert} e^{i(p_1+p_2)(x-y)/z} \frac{p_{12}}{p_{12} + g} ~~~~~. \label{eq:Pitilde}
\end{eqnarray}
Note that the factor $1 + \lambda_0 B(p_{12}^2)$ on the first line is nothing but the wavefunction renormalization $Z_O$ of the composite operator $O = \vp^2$ in Eq.~(\ref{eq:Z_O}) with the renormalization scale $\mu$ replaced by $p_{12}$,
as expected from the construction (\ref{eq:Pi}).
We use this expression to discuss the IR limit later.
Now we start with the UV limit.

\subsection{UV limit}
Since $S(p,z)=1+\cl{O}(z^2)$ in the UV limit ($z\to 0$), Eq.~(\ref{eq:G_UV}) leads to  
\begin{eqnarray}
\Pi_\lambda (x-y,z) &=& -{128 \lambda_0 z  \over 3} \left [\Omega(x-y) + \cl{O}(z^2) \right], 
\end{eqnarray}
where 
\eqn{ 
    \Omega(x) = & \int_{p_1,p_2} {e^{i(p_1+p_2)x }\over p_1^2p_2^2} {1 \over 8 p_{12} +\lambda_0 } ~.
    \label{eq:Omega}
}
The explicit form of $\Omega$ is given in Eq.~(\ref{eq:Omega-explicit}).
In total, we obtain
\begin{eqnarray}
    \Pi(X,y) &\simeq &{64 z\over 3} {1\over N}\left[ {1\over 8\pi^2}{1\over z^2+(x-y)^2}-2 \lambda_0  \Omega(x-y) \right] \label{eq:PiUV}
\end{eqnarray}
as $z\to 0$, where the second term represents a correction due to the non-zero coupling $\lambda_0$.

Therefore, even in the presence of the interaction, we see
\begin{equation}
   \Pi(X,y) \simeq  z^{\Delta_{\varphi^2}^{\rm UV}}\times {64\over 3}\left\langle \varphi^a(x)\varphi^a(x)\varphi^b(y)\varphi^b(y) \right\rangle_c + \cl{O}(z^3), \quad z\to 0 
    \label{eq:Bulk-to-boundary_UV}
\end{equation}
where the connected correlation function of $\varphi^2$ is obtained in Eq.~(\ref{eq:scalar}), and
$\Delta_{\varphi^2}^{\rm UV}=1$ corresponds to its conformal dimension at the asymptotic free UV fixed point of the boundary theory.
Moreover, as shown in appendix~\ref{app:composite}, we see 
\begin{eqnarray}
\left\langle \varphi^a(x)\varphi^a(x)\varphi^b(y)\varphi^b(y) \right\rangle_c \propto
\left\{
\begin{array}{ll}
\vert x-y\vert^{-2}, &\  \vert x-y \vert \ll \dfrac{1}{\lambda_0} \quad  \mbox{(UV)},\\
\\
\vert x-y\vert^{-4}, &\  \vert x-y \vert \gg \dfrac{1}{\lambda_0} \quad  \mbox{(IR)}.\\
\end{array}
\right.
\end{eqnarray}

\subsection{IR limit}
Let us expand the factor $p_{12}/(p_{12}+g)$ in Eq.~(\ref{eq:Pitilde}) for $g \propto z \to \infty$. Then we get
\begin{eqnarray}
\Pi(x-y,z) =\sum_{n=1}^\infty \Pi_{n}(x-y,z),
\end{eqnarray}
where
\begin{eqnarray}
    \Pi_{n}(x,z)&:=&{2\over N \gamma_0(z) z^2}{(-1)^{n+1}\over g^n}
    \int_{p_1,p_2}{K_1(p_1) K_2(p_2) \over p_1 p_2} e^{i(p_1+p_2)\cdot x/z} p_{12}^n.
\end{eqnarray}
The leading contribution  is the $n = 1$ term, which can be further expanded  as
\begin{eqnarray}
     \Pi_{1}(x,z) &=& {1\over N z^2} {512\over 3\pi^4 \lambda_0 }\sum_{k=0}^\infty {(k+1) (-1)^k A_k\over (2k+3)!}\left( {x^2\over z^2}\right)^k,
    \label{eq:Pi1IR}
\end{eqnarray}
where
\begin{eqnarray}
    A_k &:=&\int_0^\infty dp_1\, \int_0^{p_1}dp_2\, K_1(p_1) K_1(p_2) \left[(p_1+p_2)^{2k+3} -(p_1-p_2)^{2k+3}\right] ~.
    \label{eq:Ak}
\end{eqnarray}
This integral converges, and especially, we find $A_0 = 3 \pi^2 /2$.

Thus, at large $z$, the bulk-to-boundary propagator behaves as
\begin{eqnarray}
    \Pi(X,y) = {1\over z^2} {256 A_0\over 9\pi^4 \lambda_0 N } \left[ 1 +\cl{O}\left(z^{-1}\right) \right] = {1\over z^2} {128 \over 3\pi^2 \lambda_0 N }  \left[ 1 +\cl{O}\left(z^{-1}\right) \right],
    \label{eq:Bulk-to-boundary_IR}
\end{eqnarray}
where the NLO contribution $\propto z^{-3}$ in the large $z$ comes from $\Pi_{2}(x-y,z)$.
This behavior is consistent with the LO behavior of the bulk-to-boundary propagator for the scalar field with the conformal dimension $\Delta_{\varphi^2}^{\rm IR}$ in the presence of conformal symmetry, which is 
\begin{equation}
    \left[{z\over z^2 + (x-y)^2}\right]^{\Delta_{\varphi^2}^{\rm IR}} \sim z^{-\Delta_{\varphi^2}^{\rm IR}}, \quad z\to\infty .
\end{equation}
The $z^{-\Delta_{\varphi^2}^{\rm IR}}$ behavior also satisfies the EOM of the free scalar field at $z\to\infty$, whose mass $m$ is given by $m^2 \propto \Delta_{\varphi^2}^{\rm IR}(\Delta_{\varphi^2}^{\rm IR}-d)$.
Therefore the $z$ dependence of the bulk-to-boundary scalar propagator at $z\to\infty$ in Eq.~(\ref{eq:Bulk-to-boundary_IR}) correctly reproduces $\Delta_{\varphi^2}^{\rm IR} =2$ (the conformal dimension of $\varphi^2$) at the Wilson-Fisher IR fixed point.

In this section, we have shown that $z\to 0$ and $z\to \infty$ behaviors of the bulk-to-boundary propagator constructed by the conformal smearing correspond to conformal dimensions of $\varphi^2$ 
at the asymptotic free UV and Wilson-Fisher IR fixed points, respectively,  
in the interacting theory. 
Note that this property is robust in the sense that it holds even for a general smearing function $S(x,z)$ rather than the conformal smearing, as discussed in the appendix \ref{app:general}.

\section{Conclusion}
\begin{figure}[t]
\begin{center}
\includegraphics[width=12cm]{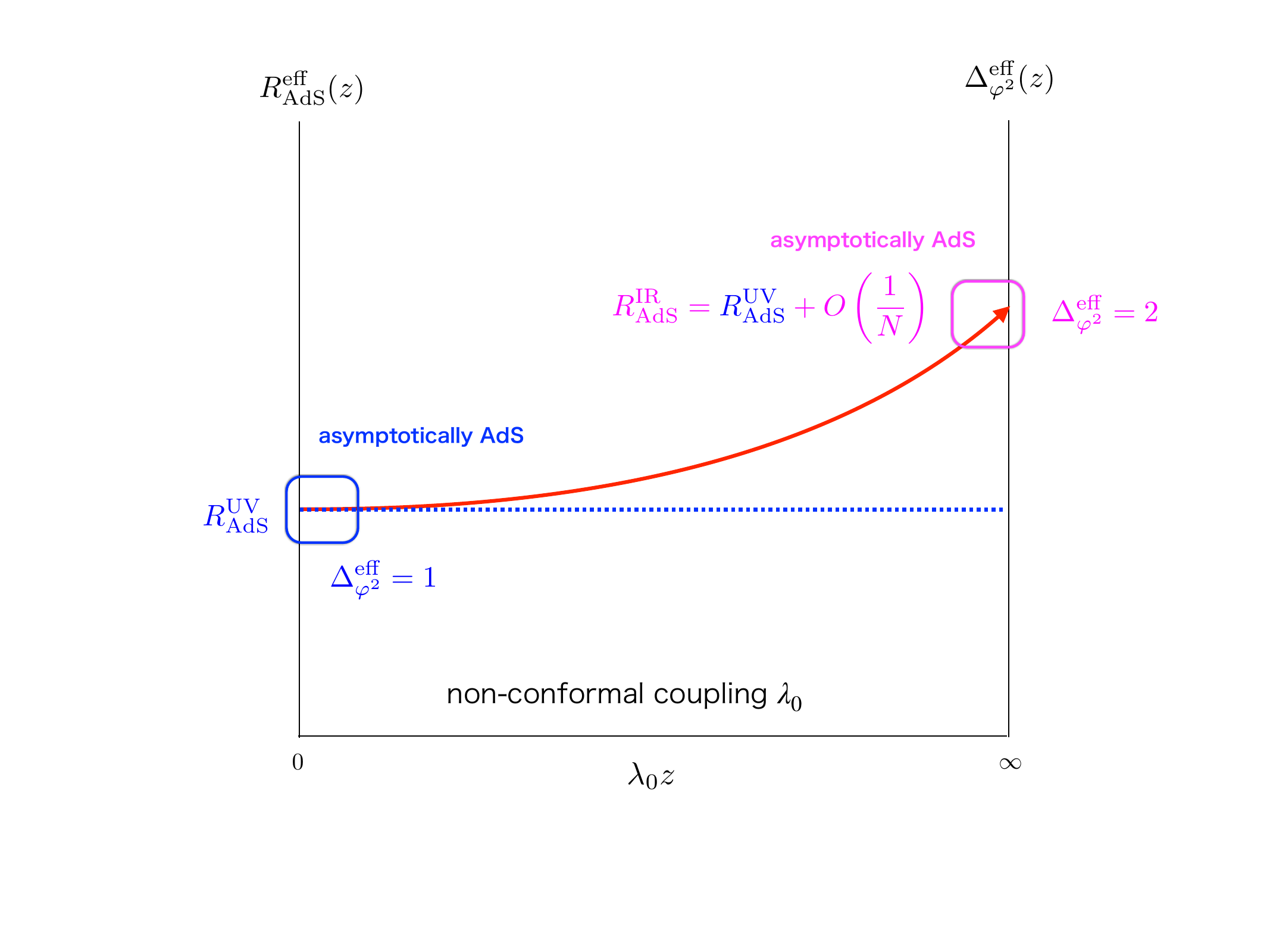}
\end{center}
\vskip -1.5cm
\caption{A schematic figure for the (effective) AdS radius ($R_{\rm AdS}^{\rm eff}(z)$) and the (effective) conformal dimension of $\varphi^2$ ($\Delta_{\varphi^2}^{\rm eff}(z)$) as a function of $\lambda_0 z$.
}
\label{fig:bulk_space}
\end{figure}

In this paper, we have investigated the bulk space dual to $O(N)$ invariant critical $\varphi^4$ model in 3-dimensions combining the conformal smearing with the large $N$ expansion, and obtained the following results, which are also schematically summarized in Fig.~\ref{fig:bulk_space}.

The metric in the bulk space at the NLO is given by Eq.~\eqref{eq:metric_NLO}, which describes the asymptotic AdS space both UV~($z\to 0$) limit in Eq.~\eqref{eq:metric_UV}
and IR~($z\to \infty$) limit in Eq.~\eqref{eq:metric_IR}.
Moreover, the AdS radii satisfy $R_{\rm AdS}^{\rm IR} -R_{\rm AdS}^{\rm UV} = \cl{O}(1/N) > 0 $, which 
reflects the fact that the number of the degrees of freedom decreases from UV to IR by the renormalization group.
This result, however, is opposite to the prediction by the F-theorem~\cite{Freedman:1999gp, Myers:2010xs, Myers:2010tj, Jafferis:2011zi, Klebanov:2011gs,Pufu:2016zxm}.
To understand this discrepancy is left to future investigation.

The bulk-to-boundary propagator in Eq.~\eqref{eq:Pi} at the LO encodes $\Delta_{\varphi^2}$ (the conformal dimension of the composite scalar operator $\varphi^2$) in its $z$ dependence at UV
as Eq.~\eqref{eq:Bulk-to-boundary_UV} and at IR as Eq.~\eqref{eq:Bulk-to-boundary_IR}, corresponding to
$\Delta_{\varphi^2}=1$ at the asymptotic UV fixed point and $\Delta_{\varphi^2}=2$ at the Wilson-Fisher IR fixed point, respectively. Interestingly, the UV limit in Eq.~\eqref{eq:Bulk-to-boundary_UV}  reproduces an expected GKP-Witten relation for the interacting theory with non-zero $\lambda_0$,
whose $z^{\Delta_{\varphi^2}}$ behavior is controlled by $\Delta_{\varphi^2}=1$, the value at the UV fixed point, while the $\vert x-y\vert$ behavior of the two-point function for $\varphi^2$ shows
complicated behavior that $\vert x-y\vert^{-2}$ at $\lambda_0 \vert x-y\vert \ll 1$ (UV in the $O(N)$ model) ) or $\vert x-y\vert^{-4}$ at $\lambda_0 \vert x-y\vert \gg 1$ (IR in the $O(N)$ model).

\vskip 0.5cm

As the conformal smearing approach works well for the $O(N)$ invariant critical $\varphi^4$ model in 3-dimensions, 
one may use it to derive some properties of the higher spin theories in 4-dimensions~\cite{Vasiliev:1995dn}, which is expected to be dual to the $O(N)$ model in 3-dimensions.


\begin{acknowledgments}
\noindent
This work has been supported in part by KIAS Individual Grants, Grant No. 090901, and by the JSPS Grant-in-Aid for Scientific Research (No. JP22H00129).
We would like to thank the referee of this paper for pointing out a discrepancy between our result and that from the F-theorem.
\end{acknowledgments}

\bibliographystyle{JHEP}
\bibliography{BulkReconstruction}

\appendix
\section{$O(N)$ model in 3 dimensions}
\label{sec:largeN}
\subsection{Schwinger-Dyson equation}

The Schwinger-Dyson equation (SDE) for the action $S(\varphi)$ in eq.~(\ref{eq:action}) is compactly written as
\begin{equation}
\left\langle {\delta O(\varphi)\over \delta\varphi(x)}\right \rangle    
= \left\langle O(\varphi) {\delta S(\varphi) \over \delta\varphi(x)}\right \rangle 
\end{equation}
where 
\begin{equation}
\langle O(\varphi) \rangle    
:= {1\over Z}\int {\cal D}\varphi \, O(\varphi) e^{-S(\varphi)}, \quad
Z:= \int {\cal D}\varphi \,  e^{-S(\varphi)} .
\end{equation}

We define the connected part of $n$-point functions with appropriate powers of $N$ as
\begin{eqnarray}
\Gamma^{ab}(x,y) &:=& N \langle \varphi^a(x) \varphi^b(y) \rangle, \\
K^{a_1a_2\cdots a_n} &=& N^{n-1} \langle \varphi^{a_1}(x_1) \varphi^{a_2}(x_2)\cdots \varphi^{a_n}(x_n) \rangle_c
\end{eqnarray}
for $n=4,6,,\cdots$. The $O(N)$ symmetry tells us 
\begin{eqnarray}
\Gamma^{ab}(x,y) &:=& \delta^{ab} \Gamma(x-y) \\
K^{a_1a_2a_3a_4} &=&  \delta^{a_1a_2}  \delta^{a_3a_4} K(x_1,x_2;x_3,x_3) + (2\leftrightarrow 3) + (2\leftrightarrow 4), \\
K^{a_1a_2\cdots a_6}(x_1,x_2,\cdots,x_6) &=& \delta^{a_1a_2}  \delta^{a_3a_4}\delta^{a_5a_6} H(x_1,x_2;x_3,x_4;x_5.x_6) + \mbox{(14 perm.)}.
\end{eqnarray}

Taking $O=\varphi^a(x)$, the Schwinger-Dyson equation leads to
\begin{eqnarray}
\delta^{(3)}(x-y) &=& \left[-\Box + m^2 Z_m +\lambda Z_\lambda Z_\varphi \Gamma(0)\right]Z_\varphi \Gamma(x-y)\nonumber\\
&+& {\lambda Z_\lambda Z_\varphi^2\over N}\left[\left(1+{2\over N}\right)K(x,y;x,x) + 2\Gamma(0) \Gamma(x-y)\right],
\label{eq:SD_phi}
\end{eqnarray}
while the one for $O=\varphi^{a_2}(x_2)\varphi^{a_3}(x_3) \varphi^{a_4}(x_4)$ gives 
\begin{eqnarray}
0 &=& \left[-\Box + m^2 Z_m  +\lambda Z_\lambda Z_\varphi\left(1+{2\over N}\right) \Gamma(0)\right]Z^2_\varphi K(x_1,x_2;x_3,x_4)\nonumber\\
&+& \lambda Z_\lambda Z_\varphi^3\Gamma(x_1-x_2)\left[2\Gamma(x_1-x_3)\Gamma(x_1-x_4) +\left(1+{2\over N}\right)K(x_1,x_1;x_3,x_4) + {2\over N}K(x_1,x_3;x_1,x_4)\right]\nonumber \\
&+&{\lambda Z_\lambda Z_\varphi^3\over N}
\left[\left(1+{2\over N}\right)H(x_1,x_1;x_1,x_2;x_3,x_4) + {2\over N}H(x_1,x_2;x_1,x_3;x_1,x_4)\right]\nonumber \\
&+& {2\lambda Z_\lambda Z_\varphi^3\over N}
\left[\Gamma(x_1-x_3)K(x_1,x_2;x_1,x_4) + \Gamma(x_1-x_4)K(x_1,x_2;x_1,x_3)\right],
\end{eqnarray}
where eq.~\eqref{eq:SD_phi} has already been used.

\subsection{Large $N$ expansion}
\subsubsection{2-pt function at the LO}
Using the Fourier transformation of the 2-pt function that
\begin{equation}
\Gamma(x) =\int {d^3p\over (2\pi)^3}e^{i p x}\tilde \Gamma(p), \quad  \tilde \Gamma(p)=\tilde \Gamma_0(p) + {1\over N} \Gamma_1(p)+\cdots,  
\end{equation}
the 2-pt function at the LO satisfies
\begin{equation}
1= \left[p^2 + m^2 Z_m +\lambda Z_\lambda Z_\varphi \Gamma_0(0)\right]Z_\varphi \tilde \Gamma_0(p), \quad \Gamma_0(0)=\int {d^3p\over (2\pi)^3} \tilde \Gamma_0(p).
\end{equation}

We adopt the renormalization conditions for the 2-pt function that 
\begin{equation}
\left.\tilde \Gamma_0^{-1}(p)\right\vert_{p^2=\mu^2}= m^2+\mu^2, \quad
\left.{d\over dp^2}\tilde \Gamma_0^{-1}(p)\right\vert_{p^2}=1,
\end{equation}
which lead to
\begin{equation}
Z_\varphi=1, \quad
m^2Z_m=m^2 +\lambda Z_\lambda\left({m\over 4\pi}-{\Lambda\over 2\pi^2}\right),  
\end{equation}
where we use
\begin{equation}
\Gamma_0(0)=\int {d^3p\over (2\pi)^3} {1\over p^2+m^2}={\Lambda\over 2\pi^2}-{m\over 4\pi}
\end{equation}
with $\Lambda$ being the momentum cut-off and $m\ge 0$ being assumed.

\subsubsection{4-pt function at the LO}

The SDE relevant for the 4-pt function at the LO
is written as
\begin{equation}
( -\Box_1 +m^2)K_0(12;34)+\lambda Z_\lambda \Gamma_0(12) K_0(11;34) = -2\lambda Z_\lambda \Gamma_0(12) \Gamma_0(13) \Gamma_0(14),  
\label{eq:4pt_LO}
\end{equation}
where we expand $K=K_0+ K_1/N +\cdots$, and employ short-handed notations such as $\Gamma_0(ij)=\Gamma(x_i-x_j)$ and 
$K_0(ij;kl)=K(x_i,x_j;x_k,x_l)$.

Introducing the amputated 4-pt function in the momentum space as
\begin{equation}
K_0(12;34) :=\prod_{i=1}^4 \left(\int {d^3p_i\over (2\pi)^3}{e^{i p_i x_i}\over p_i^2+m^2}\right)
(2\pi)^3\delta^{(3)}\left(\sum_{i=1}^4 p_i\right)
G_0(p_1,p_2;p_3,p_4),
\end{equation}
the SDE becomes 
\begin{eqnarray}
G_0(p_1,p_2;p_3,p_4) &+&\lambda Z_\lambda 
\prod_{i=1}^2 \left(\int {d^3q_i\over (2\pi)^3}{1\over q_i^2+n^2}\right) G_0(q_1,q_2;p_3,p_4)
\delta^{(3)}\left(q_1+q_2+p_3+p_4\right)\nonumber \\
&=&-2\lambda Z_\lambda,
\end{eqnarray}
where $p_1+p_2+p_3+p_4=0$.

The solution to the above equation is given by
$G_0(p_1,p_2;p_3+p_4)=G_0(p_{12}^2)$ with $p_{12}:=p_1+p_2$ and
\begin{equation}
G_0(p^2) ={-2\lambda Z_\lambda\over 1+\lambda Z_\lambda B(p^2)},    
\end{equation}
where
\begin{equation}
B(p^2) =\int {d^3q\over (2\pi)^3} {1\over q^2+m^2}{1\over (p-q)^2+m^2} = \frac{\pi -2 \arctan (2m/|p|)}{8 \pi |p|} ,
\end{equation}
which becomes simple in the massless limit as
\begin{equation}
\lim_{m^2\to 0} B(p^2) ={1\over 8\vert p\vert} ,
\end{equation}
and thus
\begin{equation}
\lim_{m^2\to 0} G_0(p_1,p_2;p_3,p_4) =\dfrac{-2\lambda Z_\lambda} {1+\lambda Z_\lambda/ (8\vert p_{12}\vert)} ,
\end{equation}

We take the renormalization condition for the amputated 4-pt function as
\begin{equation}
\left. G(p_1,p_2;p_3,p_4)\right\vert_{p_{12}^2=\mu^2} =-2\lambda,
\end{equation}
which gives $G_0(\mu^2)=-2\lambda$ at the LO.
We thus obtain
\begin{equation}
Z_\lambda={1\over 1-\lambda B(\mu^2)}\longrightarrow \dfrac{1} {1-\lambda/ (8\mu)} , \quad m^2\to 0.
\end{equation}

Now all renormalization constants are fixed at the LO.

\subsubsection{2-pt function at the NLO}
The 2-pt function at the NLO,
necessary in the main text, is also calculated here.

At the NLO, we write
\begin{equation}
Z_\varphi = 1+{1\over N} Z_\varphi^{(1)}, \quad
Z_m = Z_m^{(0)}+{1\over N} Z_m^{(1)},
Z_\lambda=Z_\lambda^{(0)}+{1\over N}Z_\lambda^{(1)},
\end{equation}
where the LO parts have already been determined.

The SDE relevant for the 2-pt function at the NLO reads
\begin{eqnarray}
0&=&(-\Box+m^2) \Gamma_1(12) +\left[m^2( Z_m^{(1)}+Z_m^{(0)}Z_\varphi^{(1)}) +\lambda_0(\Gamma_1(0) + {Z_\lambda^{(1)}\over Z_\lambda^{(0)}}+2Z_\varphi^{(1)}\right]
\nonumber \\
&+& \lambda_0 \left[K_0(12;11)+2\Gamma_0(0)\Gamma_0(12)\right].
\end{eqnarray}
where $\lambda_0:=\lambda Z_\lambda^{(0)}$ at this order.
Using the relation obtained from eq.~\eqref{eq:4pt_LO} as
\begin{equation}
\lambda_0 \left[K_0(x,x;0x,0)+2\Gamma_0(0) \Gamma_0(x)\right] =-\left.{1\over \Gamma_0(0)}(-\Box_1+m^2)K_0(x_1,x;x,0)\right\vert_{x_1=x},    
\end{equation}
the SDE becomes
\begin{eqnarray}
(-\Box + m^2) \Gamma_1(12) &=& -\left[ m^2 \left(Z_m^{(1)}+Z_m^{(0)}Z_\varphi^{(1)}\right)+\lambda_0\left(\Gamma_1(0) +{Z_\lambda^{(1)}\over Z_\lambda^{(0)}}+2 Z_\varphi^{(1)}\right)\right] \nonumber \\
&+&\left.{-\Box_1+m^2\over \Gamma_0(0)} K_0(x_1,x;x,0) \right.\vert_{x_1=x} ,
\end{eqnarray}
whose last term is further evaluated as
\begin{eqnarray}
&&\left.{-\Box_1+m^2\over \Gamma(0)} K_0(x_1,x;x,0) \right\vert_{x_1=x}= \int{d^3p\over (2\pi)^3}{e^{ipx}\over p^2+m^2}\int{d^3Q\over (2\pi)^3}{G_0(Q)\over (Q-p)^2+m^2}  \nonumber \\
&=&
\int{d^3p\over (2\pi)^3}{e^{ipx}\over p^2+m^2} 
\int{d^3Q\over (2\pi)^3}{1\over (Q-p)^2+m^2}
{-2\lambda_0\over 1+\lambda_0 B(Q^2)}.
\end{eqnarray}

Using the expression in the momentum space as
\begin{equation}
\Gamma_1(x)=\int {d^3p\over (2\pi)^3}{X(p^2)\over (p^2+m^2)^2}e^{i px}, 
\label{eq:Gamma1x}
\end{equation}
the SDE leads to
\begin{equation}
X(p^2) = -\left[m^2\left(Z_m^{(1)}+Z_m^{(0)}Z_\varphi^{(1)}\right)+\lambda_0\left(\Gamma_1(0) +{Z_\lambda^{(1)}\over Z_\lambda^{(0)}}+2 Z_\varphi^{(1)}+2 Y(p^2)\right)\right],
\label{eq:X}
\end{equation}
where by definition
\begin{eqnarray}
\Gamma_1(0) &=& \int {d^3p\over (2\pi)^3}{X(p^2)\over (p^2+m^2)^2}, 
\label{eq:Gamma10}
\end{eqnarray}
while
\begin{eqnarray}
Y(p^2) &=& \int {d^3Q\over (2\pi)^3}{1\over (Q-p)^2+m^2}{1\over 1+\lambda_0 B(Q^2)}.    
\end{eqnarray}

Inserting eq.~\eqref{eq:X} into eq.~\eqref{eq:Gamma10}, we obtain  
\begin{eqnarray}
\Gamma_1(0) &=& -\frac{m^2\left(Z_m^{(1)}+Z_m^{(0)}Z_\varphi^{(1)}\right)+\lambda_0\left(\Gamma_1(0) +{Z_\lambda^{(1)}\over Z_\lambda^{(0)}}+2 Z_\varphi^{(1)}+2 Y(p^2)\right)} {1+\lambda_0 I(m^2)} ,  
\end{eqnarray}
where
\begin{equation}
I(m^2) :=\int{d^3p\over (2\pi)^3}{1\over (p^2+m^2)^2}= -{d\over dm^2}\Gamma_0(0) ={1\over 8\pi}{1\over \sqrt{m^2}}.   
\end{equation}

The 2-pt function at the NLO in the momentum space is expressed as
\begin{eqnarray}
Z_\varphi \tilde \Gamma(p) &=& \left(1+{1\over N} Z_\varphi^{(1)}\right) \tilde \Gamma_0(p) +{1\over N} \tilde \Gamma_1(p) 
=\frac{1+{1\over N}Z_\varphi^{(1)}}{p^2+ m^2 -{1\over N}X(p^2)},
\end{eqnarray}
and thus the renormalization condition at $p^2=\mu^2$ reads
\begin{eqnarray}
 \left(1-{1\over N} Z_\varphi^{(1)}\right) (\mu^2+m^2) -  {1\over N}X(p^2) &=&\mu^2+m^2, \\
 1-{1\over N} Z_\varphi^{(1)} -\left.{1\over N}{d\over dp^2}X(p^2)\right\vert_{p^2=\mu^2} = 1,
\end{eqnarray}
which leads to relations among renormalization constants at the NLO as
\begin{eqnarray}
m^2 \left( Z_m^{(1)} + Z_m^{(0)} Z_\varphi^{(1)}\right) &=& Z_\varphi^{(1)}(\mu^2+m^2) -\lambda_0 \left(\Gamma_1(0) + {Z_\lambda^{(1)}\over Z_\lambda^{(0)}} + 2 Z_\varphi^{(1)} + 2 Y(\mu^2)\right),~~~~ \\
Z_\varphi^{(1)} &=& \left. 2\lambda_0 {d\over dp^2} Y(p^2)\right\vert_{p^2=\mu^2}  .         
\end{eqnarray}

We then finally obtain
\begin{equation}
X(p^2) = -2\lambda_0 Y_r(p^2) -Z_\varphi^{(1)}(\mu^2+m^2), \quad Y_r(p^2):= Y(p^2) -Y(\mu^2),    
\end{equation}
which is UV-finite, thanks to the subtraction.

In the main text, we consider the case with $m^2=\mu^2=0$, which leads to 
\begin{eqnarray}
X(p^2) = \int {d^3Q\over (2\pi)^3} \left[{1\over (Q-P)^2}-{1\over Q^2}\right]{-2\lambda_0\over 1+\lambda_0 B(Q^2)}= {2\lambda_0 p\over (2\pi)^2} L\left( {\lambda_0\over 8 p}\right),
\end{eqnarray}
where
\begin{eqnarray}
    L(x) &=& {x^2\over 2}\left[ \ln^2 {x+1\over x} -\left(\ln {x-1\over x}-2i\pi\right)\ln {x-1\over x} +4{\ln x +1\over x} \right]\nonumber \\
    &+& x^2\left[{\rm Li}_2\left({x\over x+1}\right)-{\rm Li}_2\left({x\over x-1}\right)\right]
\end{eqnarray}
with 
\begin{equation}
    {\rm Li}_2(x) := -\int_0^x dt\, {\ln (1-t)\over t}.
\end{equation}
In two limits ($x\to 0$ or $x\to\infty$), we have
\begin{equation}
    L(x) \longrightarrow \left\{
    \begin{array}{cc}
  \displaystyle   2(1+\ln x) x -{\pi^2\over 2}x^2 +\cdots, & x\to 0,  \\
  \\
    \displaystyle -{2\over 9x}( 1+ 3\ln x) +\cdots,    & x\to \infty.
    \end{array}
    \right. \label{L_expanded}
\end{equation}

\subsection{Renormalization Group analysis in the massless case at the LO}
\subsubsection{Beta function}
We define a dimension less coupling constant as $g_R:=\lambda/\mu$ at the LO, which is written in terms of $\lambda_0$ at $m^2 = 0$ as
\begin{equation}
g_R ={\lambda_0/\mu\over 1+\lambda_0/(8\mu)} .    
\end{equation}
Since $\lambda_0$ is $\mu$ independent, the $\beta$ function for $g_R$ can be calculated as
\begin{equation}
\beta(g_R):=\mu{d\over d\mu} g_R(\mu) = - g_R \left(1-{g_R\over 8}\right).    
\end{equation}
Therefore $g_R=0$ corresponds to an asymptotic free (UV) fixed point, while $g_R=8$ is the Wilson-Fisher (IR) fixed point.

\subsubsection{Anomalous mass dimension of the composite scalar operator}
\label{app:composite}
We calculate the anomalous dimension of the $O(N)$ invariant scalar operator, given by
\begin{equation}
O_R:=Z_O \varphi_0^a \varphi_0^a = Z_O Z_\varphi \varphi^a\varphi^a,   
\end{equation}
where $Z_O$ is the renormalization factor, from which the anomalous mass dimension of $O$ is defined as
\begin{equation}
\gamma_O:=-\mu {d\over d\mu} \ln Z_O. 
\end{equation}

At the LO where $Z_\varphi=1$, we have
\begin{equation}
    \langle  O(x_1) \varphi^{a_3}(x_3) \varphi^{a_4}(x_4)\rangle ={Z_O \delta_{a_3a_4}\over N^2} \left[2\Gamma_0(13)\Gamma_0(14) + K(_0(11;34)\right],
\end{equation}
whose Fourier transformation should be equal to the tree-level contribution that
\begin{equation}
\prod_{i=3}^4\int d^3x_i e^{i p_i x_i}\, \langle  O(x_1=0) \varphi^{a_3}(x_3) \varphi^{a_4}(x_4)\rangle\vert_{\rm tree} =   { \delta_{a_3a_4}\over N^2}  2\tilde\Gamma_0(p_3)\tilde\Gamma_0(p_4)
\end{equation}
at $p_{34}^2=\mu^2$.
Since 
\begin{equation}
  \prod_{i=3}^4\int d^3x_i e^{i p_i x_i}\,  K_0(11;34) = \tilde\Gamma_0(p_3)\tilde\Gamma_0(p_4) \int {d^3p\over (2\pi)^3}{G_0(p_{34}^2)\over [ p^2+m^2] \left[ (p_{34}-p)^2+m^2\right]},
\end{equation}
we obtain
\begin{equation}
    Z_O = \left[1+{1\over 2} G_0(\mu^2) B(\mu^2)\right]^{-1} = 1+ \lambda_0 B(\mu^2). \label{eq:Z_O}
\end{equation}

Therefore the anomalous dimension is calculated as
\begin{equation}
    \gamma_O = -{\lambda_0\over 1+\lambda_0 B(\mu^2)} \mu {d\over d\mu} B(\mu^2).
\end{equation}

In the massless case ($m^2=0$), we obtain
\begin{equation}
    \gamma_O= {g_R\over 8}  = {\lambda_0\over 8\mu + \lambda_0}= \left\{
    \begin{array}{cc}
     0    & \hspace{2em} (\mu\to \infty ),  \\
     \\
     1    & \hspace{2em} ( \mu\to 0). \\
     \end{array}\right.
\end{equation}
Therefore, the total conformal (mass) dimension of $O$ in the IR fixed point at $\mu\to 0$ becomes
\begin{equation}
    \Delta_O := 2\Delta_{\varphi_0} +\gamma_O =2, \label{eq:Delta_O-IR}
\end{equation}
where $\Delta_{\varphi_0}=1/2$ is the conformal dimension of $\varphi_0$, while $\Delta_O=1$ in the UV fixed point ($\mu\to\infty$). 

From the above result, it is straightforward to calculate the connected 2-pt function of $\varphi^2$ without $Z_O$ at the LO in the large $N$ expansion. In the massless limit, it becomes  
\begin{eqnarray}
    \langle \varphi^2(x) \varphi^2(y) \rangle _c &=&{2\over N} 
    \int_{p_1,p_2} {e^{i(p_1+p_2)\cdot (x-y)}\over p_1^2 p_2^2} {1\over 1 + \lambda_0 B(p_{12}^2) }\\   
&=&     {2\over N}\left[ {1\over 16 \pi^2 (x-y)^2}- \lambda_0 \Omega(x-y)\right],
\label{eq:scalar}
\end{eqnarray}
where $\Omega(x)$ is defined in eq.~(\ref{eq:Omega}),
whose explicit form can be written as
\eqn{ 
    \Omega(x) 
   =&  \frac{|x|^{-1}}{128 \pi^2} \qty[ \tx{Ci}(\chi) \sin (\chi) -  \tx{si}(\chi) \cos (\chi)  ] 
    \label{eq:Omega-explicit}
    }
with $\chi := |x| \lambda_0/8$ and the trigonometric integrals
\eqn{
\tx{Ci}(z) := - \int^\infty_z d t \frac{\cos t}{t} ~, ~~~ \tx{si}(z) := - \int^\infty_z d t \frac{\sin t}{t} ~.
}

In the UV limit that $|x-y| \ll 8/\lambda_0$, we get $\Omega (x-y) \approx |x-y|^{-1}/256 \pi$, and thus, the tree level contribution $\propto (x-y)^{-2}$ in (\ref{eq:scalar}) dominates the 2-pt function.

In the IR limit that $|x-y| \gg 8/\lambda_0$, on the other hand, we have
\eqn{\Omega (x-y) = \frac{(x-y)^{-2}}{16 \pi^2 \lambda_0} \qty(1  - \frac{2 (8/ \lambda_0)^2}{(x-y)^2} + \cl{O}\qty( (x-y)^{-4}) )~. } 
Therefore, the 2-pt function (\ref{eq:scalar}) behaves as
\eqn{
\langle \varphi^2(x) \varphi^2(y) \rangle _c \approx \frac{1}{N} \frac{16}{\pi^2 \lambda_0^2} \frac{1}{(x-y)^4} ~,
}
as expected from (\ref{eq:Delta_O-IR}).

\section{Metric at the NLO}
The metric at the NLO is given in eq.~\eqref{eq:metric_NLO} in the main text as
\begin{eqnarray}
    g_{\mu\nu}^{\rm NLO}(z) &=& g_{\mu\nu}^{\rm LO}(z) \left[1 +{1\over N} G_s(z)\right], \quad
    g_{zz}^{\rm NLO}(z) = g_{zz}^{\rm LO}(z) \left[1 +{1\over  N}G_\sigma(z)\right],
\end{eqnarray}
where
\begin{eqnarray}
G_s(z)&:=& {F_1(z)\over F_0(z)} -{\gamma_1(z)\over \gamma_0(z)} ={32 g \over \pi^4}\left[ {64\over 45} F_{11}^3(g) - F_{11}^1(g)\right], 
\label{eq:Gs}\\
G_\sigma(z) &:=&{1\over 5} \left( {9H_1(z)\over H_0(z)} -{\gamma_1(z)\over \gamma_0(z)}-{8G_1(z)\over   G_0(z)} \right)={32 g \over 5\pi^4}\left[{64\over 3} \left\{F_{00}^3(g)-F_{01}^2(g)\right\}-F_{11}^1(g)\right] ,~~~~~~
\label{eq:Gr}
\end{eqnarray}
with $g:= \lambda_0 z/ 8$ and
\begin{eqnarray}
    F_{ij}^n(g) := \int_0^\infty dp\, p^n K_i(p) K_j(p) L\left({g\over p}\right).
    \label{eq:Fnij}
\end{eqnarray}
Analytic expressions of $F_{ij}^n(g)$ are given in the appendix \ref{sec:integral} in two limits that
$g\to 0$ (UV) or $\to\infty$ (IR).
\section{Momentum integrals}
\label{sec:integral}
Here we present several momentum integrals used in the main text.

\subsection{Integral formula for Bessel functions}
We present useful formulas for integrals of Bessel functions.\\

\noindent
(1) 6.576-4 in \cite{Gradshteyn:1943cpj}:
\begin{eqnarray}
    \int_0^\infty dx\, x^{-\lambda} K_\mu(ax) K_\nu(bx) &=& {2^{-2-\lambda}a^{-\nu+\lambda-1} b^\nu\over \Gamma(1-\lambda)}\prod_{\pm} \Gamma\left({1-\lambda\pm \mu +\nu\over 2}\right)\Gamma\left({1-\lambda\pm \mu -\nu\over 2}\right)\nonumber \\
    &\times& {}_2F_1\left({1-\lambda+ \mu +\nu\over 2},{1-\lambda- \mu +\nu\over 2};1-\lambda;1-{b^2\over a^2}\right)
    \label{eq:Bessel1}
\end{eqnarray}
for $\Re (a+b) > 0$, $\Re\, \lambda < \vert\Re\, \mu\vert - \vert\Re\, \nu\vert$. \\

\noindent
(2) 6.671-5 in \cite{Gradshteyn:1943cpj}:
\begin{eqnarray}
    \int_0^\infty dx\, K_\nu(ax) \sin(bx) &=&{\pi\over 4} {a^{-\nu}\over \sin(\nu\pi/2)}{\left[ (\sqrt{b^2+a^2}+b)^\nu -(\sqrt{b^2+a^2}-b)^\nu \right]\over \sqrt{a^2+b^2}}
    \label{eq:Bessel2}
\end{eqnarray}
for $\Re\, a > 0$, $b > 0$, $\vert \Re\, \nu\, \vert < 2$, $\nu\not= 0$. 

\subsection{$F_{ij}^n$ in two limit}
We split $F_{ij}^n(g)$ defined in eq.~(\ref{eq:Fnij}) into two part as
$F_{ij}^n(g)= F_{ij}^{n,a} (g) + F_{ij}^{n,b} (g)$, where
\begin{eqnarray}
   F_{ij}^{n,a} (g) &:=& \int_0^g dp\, p^n K_i(p) K_j(p) L(g/p), \quad
    F_{ij}^{n,b} (g) := \int_g^\infty dp\, p^n K_i(p) K_j(p) L(g/p),~~
\end{eqnarray}
and evaluate it in UV and IR limits.

\subsubsection{IR limit}
We first consider the IR ($g\to\infty$) limit.

Using an expansion in terms of $1/g$ at $g=\infty$ as
\begin{eqnarray}
   \int_0^g dx\, f(x) \simeq \int_0^\infty dx\, f(x)  +{1\over g} \lim_{g\to\infty} f(g) {dg\over d(1/g)}+\cdots,
\end{eqnarray}
we have 
\begin{eqnarray}
 F_{ij}^{n,a}(g) &\simeq&   \int_0^\infty dp\, p^n K_i(p) K_j(p) L(g/p)  -{1\over g} \lim_{g\to\infty} g^{n+2}K_i(g) K_j(g) L(1)+\cdots\nonumber \\
 &\simeq& -{2\over 9 g}\int_0^\infty dp p^{n+1} K_i(p) K_j(p) \left(1+3\ln g/p \right)\nonumber \\
 &=& -{1\over 3g}\left[ \left({2\over 3} +\ln g^2\right) C_{ij}^{n+1} - L_{ij}^{n+1} \right],
\end{eqnarray}
where the second term in the first line vanishes exponentially, the asymptotic behavior of $L(x)$ in (\ref{L_expanded}) is used to obtain the second line, and constants in the third line are defined as
\begin{eqnarray}
    C_{ij}^n :=\int_0^\infty dp\, p^n K_i(p) K_j(p), \quad
    L_{ij}^n :=\int_0^\infty dp\,  \ln p^2\, p^n K_i(p) K_j(p) .
    \label{eq:CLijn}
\end{eqnarray}

On the other hand, $F_{ij}^{n,b}$ is evaluated by the steepest-descent method after a change of variable $p= g y$ as
\begin{eqnarray}
    F_{ij}^{n,b}(g) &=& g^{n+1}\int_1^\infty dy\, y^n K_i(gy) K_j(gy) L(1/y) \simeq g^{n+1} y_0^n K_i(g y_0) K_j(gy_0) L(1/y_0),
\end{eqnarray}
which vanishes exponentially as $g\to\infty$, where 
$1\le y_0<\infty $ is a point which gives the largest contribution to the integral. 
The point $y_0$ is either given as a solution to $S^\prime(y)=0$, where 
\begin{equation}
 S(y) := \ln \left[ y^n K_i(gy) K_j(gy) L(1/y)\right],
\end{equation}
or $y_0=1$ if no solution exists.

In total we obtain
\begin{equation}
    F_{ij}^n(g) \simeq -{1\over 3g}\left[ \left({2\over 3} +\ln g^2\right) C_{ij}^{n+1} - L_{ij}^{n+1} \right]
\end{equation}
in the IR ($g\to\infty$) limit.

\subsubsection{UV limit}
In the UV ($g\to 0$) limit, $F_{ij}^{n,a}$ is evaluated with $p=g y$ as
\begin{eqnarray}
    F_{ij}^{n,a}(g) = g^{n+1} \int_0^1 dy\, y^n K_i(gy) K_j(gy) L(1/y) 
    \simeq c_i c_j g \int_0^1 dy\, (g y)^{n-i-j} L(1/y), 
\end{eqnarray}
where $c_i$ is given in the small $x$ expansion of the Bessel function as $K_i(x) = c_i x^{-i} + \cdots$ with
$x^{-i}$ read as $\ln x$ for $i=0$. For example, $c_0=-1$, $c_1=1$, and $c_2=2$.
The $y$ integral is convergent for $n-i-j > -2$ since $L(1/y) \sim y\ln y$ for small $y$ as seen in (\ref{L_expanded}).

For $n-i-j > 0$, $F_{ij}^{n,b}$ is evaluated straightforwardly as
\begin{eqnarray}
    F_{ij}^{n,b}(g) &\simeq& 
     2g \int_0^\infty dp\, p^{n-1} K_i(p) K_j(p) \left[1+\ln(g/p)\right]
\end{eqnarray}

For $n=i=j=1$, we calculate $F_{ij}^{n,b}$ as
\begin{eqnarray}
    F_{11}^{1,b} = \int_g^\infty dp\, p\left[ K_1^2(p) -{c_1^2\over p^2}\right] L(g/p) +
c_1^2\int_1^\infty dy\, {L(1/y) \over y},
\end{eqnarray}
where the second term is a $g$-independent constant, while the first term can be evaluated using the UV limit of $L(x)$ in (\ref{L_expanded}) by 
\begin{eqnarray}
  \int_g^\infty dp\, p\left[ K_1^2(p) -{c_1^2\over p^2}\right] L(g/p)   \simeq 2g \int_0^\infty dp\, \left[ K_1^2(p) -{c_1^2\over p^2}\right] \left( 1+\ln(g/p) \right).
\end{eqnarray}

In total, we have
\begin{eqnarray}
    F_{ij}^n(g) \simeq -g \left[(2+\ln g^2) C_{ij}^{n-1} - L_{ij}^{n-1} \right] - g^{n+1-i-j} D_{ij}^n 
    \label{eq:Fijn_UV}
\end{eqnarray}
for $n-i-j > 0$, where $C_{ij}^n$ and $L_{ij}^n$ are already given in eq.~(\ref{eq:CLijn}), and 
\begin{equation}
    D^n_{ij}:= c_i c_j \int_0^1 dy\, y^{n-i-j} L(1/y), 
\end{equation}
while for $n=i=j=1$ one should replace $C_{11}^{0}$ and $L_{11}^0$ with 
\begin{eqnarray}
\tilde C_{11}^{0} :=\int_0^\infty dp\, \left[ K_1^2(p) -{c_1\over p^2}\right], \quad
\tilde L_{11}^{0} :=\int_0^\infty dp\, \left[ K_1^2(p) -{c_1\over p^2}\right]\ln p^2,
\end{eqnarray}
respectively, and
\begin{equation}
    D_{11}^{1} := c_1^2 \int_0^\infty dy  y^{-1} L(1/y) .
    \label{eq:D110}
\end{equation}

\subsection{Some calculations in the IR limit}
In the IR limit, we have a universal formula as
\begin{eqnarray}
{X_1(z)\over X_0(z)} &\simeq&-{4\over 3\pi^2} \left[{2\over 3} +\ln g^2 -C_X\right], \quad X=\gamma, F,G,H, 
\end{eqnarray}
where
\begin{eqnarray}
C_\gamma &=& {32\over 3\pi^2} L^2_{11}, \quad
C_F ={512\over 45\pi^2} L^4_{11},   \nonumber \\
C_G &=& {64\over 3\pi^2} L^3_{01} = C_\gamma+2, \quad
C_H = {512\over 27\pi^2} L^4_{00} = -{5\over 3} C_F+{8\over 3} C_\gamma +{64\over 9}. 
\label{eq:CX}
\end{eqnarray}

By combining these, we have
\begin{eqnarray}
    G_s(z) &\simeq& -{4\over 3\pi^2} C_{\rm IR}, \quad
    G_\sigma(z) \simeq {4\over 3\pi^2} \left({48\over 5} + 3 C_{\rm IR}\right), 
    \label{eq:GssigIR}
\end{eqnarray}
where
\begin{equation}
    C_{\rm IR} := C_\gamma - C_F ={512\over 45\pi^2}\left({15\over 16} L_{11}^2 -L^4_{11} \right) .
\end{equation}
Importantly, $\ln g^2$ terms are cancelled in $G_s(z)$ and $G_\sigma(z)$, so that the metric describes the AdS space in the IR limit.

\subsection{$\Pi_{\lambda,1}(x,z)$ in the IR limit}
We here evaluate $\Pi_{\lambda,1} (x,z)$ in the IR limit.
Integrating the angle between a vector $p_1+p_2$ and a vector $x$, we have
\begin{eqnarray}
  \Pi_{\lambda,1} (x,z) &=& {4\over \gamma_0(z) z^2 g (2\pi)^4 r } \int_0^\infty p_1 dp_1\, p_2 dp_2\, 
  K_1(p_1) K_2(p_2) \int_0^\pi \sin\theta d\theta \sin ( p_{12} r ),~~~
\end{eqnarray}
where $r:= \vert x\vert/z$ and $\theta$ is an angle between $p_1$ and $p_2$, so that
$p_{12}=\sqrt{p_1^2+p_2^2+2p_1 p_2 \cos\theta}$.

By expanding $\sin (p_{12} r) $, the $\theta$ integral is performed as
\begin{eqnarray}
  &&  \sum_{k=0}^\infty {(-1)^k r^{2k+1}\over (2k+1)!} \int^1_{-1} da\, (p_1^2+p_2^2 + 2 p_1 p_2 a)^{k+1/2} \nonumber \\
  &=&  \sum_{k=0}^\infty {(-1)^k r^{2k+1}\over(2k+3) p_1 p_2 (2k+1)!}
  \left[ (p_1+p_2)^{2k+3} - \vert p_1-p_2\vert^{2k+3}\right], 
\end{eqnarray}
which leads to eq.~(\ref{eq:Pi1IR}) with eq.~(\ref{eq:Ak}).

\section{Bulk-to-boundary propagator by a general smearing}
\label{app:general}
In this appendix, we investigate the UV and the IR behaviors for  the bulk-to-boundary propagator of the composite scalar operator using a general smearing function $S(\vec p,z)$, 
which satisfies  $S(\vec p,0)=1$
since $\phi^a(x,z=0)= \varphi^a(x)$ by construction.
In order to keep the rotational symmetry at the boundary and to avoid the introduction of extra dimensionful parameters, we assume  $S(\vec p,z) = S(pz)$ with $p:=\vert \vec p\vert$.

The bulk-to-boundary is compactly written in terms of $S(pz)$ as
\begin{eqnarray}
    \Pi(X,y) &=& {2 \over N\gamma_0(z) }\int_{p_1,p_2} e^{i (p_1+p_2)\cdot (x-y)} {S(p_1 z) S(p_2 z)\over p_1^2 p_2^2}{1\over 1+\lambda_0 B(p_{12}^2)} \\
    &=&{2 \over N\gamma_0(z) z^2 } \int_{p_1,p_2} e^{i (p_1+p_2)\cdot (x-y)/z} {S(p_1 ) S(p_2 )\over p_1^2 p_2^2}{p_{12}\over p_{12}+g} ,
\end{eqnarray}
where
\begin{eqnarray}
    \gamma_0(z) :=\int_{p}{ S^2(pz)\over p^2} = {c\over z}, \quad c:= \int_{p}{ S^2(p)\over p^2}.
\end{eqnarray}

In the $z\to 0$ limit, the first expression leads to
\begin{eqnarray}
 \lim_{z\to 0}  \Pi(X,y) &=&  {2z \over N c } \int_{p_1,p_2} {e^{i (p_1+p_2)\cdot (x-y)}\over p_1^2 p_2^2}{1\over 1+\lambda_0 B(p_{12}^2)}
 = {z\over c} \langle \varphi^2(x)\varphi^2(y)\rangle_c .
\end{eqnarray}

In the $z\to\infty$ limit, on the other hand, the second expression gives
\begin{eqnarray}
 \lim_{z\to \infty}  \Pi(X,y) &=& {1\over z^2} {16 \over N c  \lambda_0 } \int_{p_1,p_2} S(p_1) S(p_2) {p_{12}\over p_1^2 p_2^2} \sim {1\over z^2},
\end{eqnarray}
which leads to $\Delta_{\varphi^2}=2$.


\end{document}